\documentclass[aps,prl,amsmath,amssymb,lengthcheck,showpacs,superscriptaddress]{revtex4-1}
\usepackage{graphicx}
\usepackage{color}

\begin{document}

\title{Spatial structure of quasi-localized vibrations in nearly jammed amorphous solids}

\author{Masanari Shimada}
\email{masanari-shimada444@g.ecc.u-tokyo.ac.jp}
\affiliation{Graduate School of Arts and Sciences, The University of Tokyo, Tokyo 153-8902, Japan}
\author{Hideyuki Mizuno}
\affiliation{Graduate School of Arts and Sciences, The University of Tokyo, Tokyo 153-8902, Japan}
\author{Matthieu Wyart}
\affiliation{Institute of Physics, EPFL, CH-1015 Lausanne, Switzerland}
\author{Atsushi Ikeda}
\affiliation{Graduate School of Arts and Sciences, The University of Tokyo, Tokyo 153-8902, Japan}

\date{\today}

\begin{abstract}
The low-temperature properties of amorphous solids are widely believed to be controlled by low-frequency quasi-localized modes.
What governs their spatial structure and density is however debated. We study these questions numerically in very large systems
as the jamming transition is approached and the pressure $p$ vanishes.  We find that these modes consist of an unstable core
in which particles undergo the buckling motions and decrease the energy, and a stable far-field component which increases the energy and prevents the buckling of the core. The size of the core diverges as $p^{-1/4}$ and its characteristic volume as $p^{-1/2}$. These features are precisely those of the anomalous modes known to cause the Boson peak in the vibrational spectrum of weakly-coordinated materials. From this correspondence we deduce that the density of quasi-localized modes 
must go as $g_{\mathrm{loc}}(\omega)\sim \omega^4/p^2$, in agreement with previous observations. Our analysis thus  unravels the nature of  quasi-localized modes in a class of amorphous materials.
\end{abstract}

\maketitle

{\it Introduction}. 
The low-temperature $T\lesssim1$K properties of amorphous solids are universal, and markedly different from those of crystals ~\cite{Zeller1971Thermal,Phillips1981Amorphous}.
Their specific heat increases linearly with $T$ and their thermal conductivity increases as $T^2$~\cite{Zeller1971Thermal,Phillips1981Amorphous}.
To explain these observations, Anderson et al.~\cite{Anderson1972Anomalous} and Phillips~\cite{Phillips1972} proposed the famous  two-level systems model, that was later on extended to the soft potential model~\cite{Karpov1983Theory,Buchenau1991Anharmonic,Buchenau1992Interaction}.
This theory postulates that  amorphous solids have low-frequency quasi-localized vibrational modes in addition to phonons, which can cause double-well structures in the energy landscape.
The universal properties of amorphous solids can then be explained in terms of the quantum tunneling of these two-level systems and their interactions with phonons. 

However, the current theory is phenomenological and does not specify the nature of these localized modes, which remains a matter of debate \cite{Yu88,Gurevich03}. 
This state of affaire  led to a considerable effort to characterize quasi-localized modes numerically. 
Schober and Laird detected them in molecular-dynamics (MD) simulations in a model amorphous solid composed of the soft spheres~\cite{Schober1991}, later extended to Lennard-Jones glasses~\cite{Mazzacurati1996Low,Schober2004Size}, vitreous silica~\cite{Taraskin1999Anharmonicity}, amorphous silicon~\cite{Beltukov_2016}, and weakly-jammed solids~\cite{Xu2010Anharmonic}. 
It was found that these modes (i) have strong anharmonicity~\cite{Taraskin1999Anharmonicity,Xu2010Anharmonic}, in consistence with the assumption of the soft potential model~\cite{Karpov1983Theory,Buchenau1991Anharmonic,Buchenau1992Interaction}. (ii) Display a vibrational density of states (vDOS)  $g_{\mathrm{loc}}(\omega)$ that follows a power-law $g_{\mathrm{loc}}(\omega) \propto \omega^4$~\cite{Baity-Jesi2015,Lerner2016Statistics,Gartner2016Nonlinear,Mizuno2017Continuum,Shimada2018Anomalous} where $\omega$ is the frequency, in agreement with previous arguments for disordered bosonic systems \cite{Gurarie03,Gurevich03}. (iii) Decay  algebraically in space as long as they are not hybridized with phonon ~\cite{Lerner2016Statistics,Bouchbinder2018Universal}.
This decay is rapid enough for their participation ratio to scale as $1/N$ as for truly localized modes, where $N$ is the number of particles.  
(iv) Are suppressed if pre-stress is removed \cite{Mizuno2017Continuum,Lerner2017Frustration}. (v) Play an important role in mechanical failure under load~\cite{Maloney2006,Tanguy2010,Manning2011} and in the structural relaxation near the glass transition~\cite{Oligschleger1999,Widmer-Cooper2009}.
Interestingly, their characteristic frequency appears to rise rapidly with approaching that transition, in concert with a local measure of elastic stiffness~\cite{Lerner18}.
Despite these recent advances, understanding what fixes the nature and density of these modes remains a challenge.

In this letter we seek to resolve these questions by studying the spatial architecture of these modes, and how it is affected by the proximity of the jamming transition. 
The latter is reached  in finite-range interacting particles as the pressure $p$ vanishes~\cite{O'Hern2002Random,O'hern2003Jamming}.
A well-known property of the vibrational spectrum of amorphous solids,  an excess modes with respect to the Debye density of states called the Boson peak \cite{Phillips1981Amorphous},
is singular at that point. The associated modes, called "anomalous" in this context, have been characterized  in detail
~\cite{O'hern2003Jamming,Silbert2005Vibrations,Wyart2005Geometric,Wyart2005Effects,Silbert2009Normal,Hecke2010,Wyart2010Scaling,Ikeda2013,DeGiuli2014Effects,Charbonneau2016,Mizuno2016Elastic,Yan2016On}. By considering very large systems, we can study localized soft modes even close to jamming. 
We find that these modes consist of an unstable core
in which particles undergo the buckling motions and decrease the energy, and a stable far-field component which increases the energy and prevents the buckling of the core.
We find that the size of the core diverges as $p^{-1/4}$ and its characteristic volume as $p^{-1/2}$. All these features are precisely those of the anomalous 
modes  at the Boson peak frequency if pre-stress is removed (corresponding to removing all forces between interacting particles). Our analysis thus supports that localized soft modes are anomalous modes shifted to lower frequencies  by the destabilizing "buckling" effect of pre-stress. From this result we can immediately deduce that density of quasi-localized modes 
must go as $g_{\mathrm{loc}}(\omega)\sim \omega^4/p^2$, in agreement with previous observation~\cite{Mizuno2017Continuum}. We finally discuss how our result on the nature of localized modes generalizes to other glasses.

{\it Methods}.
We used monodisperse, three-dimensional packings of particles with mass $m$ interacting through a finite-range, harmonic potential (see Ref.~\cite{Mizuno2017Continuum} for details):
\begin{equation} \label{eq:potential}
\phi(r) = \frac{\epsilon}{2}\left(1-\frac{r}{\sigma}\right)^2H(\sigma-r), 
\end{equation}
where $\sigma$ is the particle diameter, $\epsilon$ is the characteristic energy, and $H(r)$ is the Heaviside step function.
Length, mass, and time are measured in units of $\sigma$, $m$, and $\sqrt{m\sigma^2/\epsilon}$, respectively.
The packings were generated by quenching random configurations to mechanically stable inherent structures by the FIRE algorithm~\cite{Bitzek2006Structural} and then removing the rattlers that have less than $d$ contacting neighbors.
We prepared $16$ packings for pressure $p=0.05, 0.02, 0.01, 0.005$, $15$ packings for $p=0.002$, and $8$ packings for $p=0.001$.
The system size (number of particles) is fixed at $N=1024000$, while we also analyzed $N=256000$ systems and confirmed no system size dependence in the analyzed quantities.

We next analyze the vibrational modes of these packings.
We denote the $k$-th eigenvector as $\boldsymbol{e}^k = [\boldsymbol{e}^k_1, \boldsymbol{e}^k_2,\cdots,\boldsymbol{e}^k_N]$ and its eigenvalue as $\lambda^k = (\omega^k)^2$, where $\omega^k$ is its eigenfrequency.
Note that the eigenvectors are ortho-normalized.
After removing three translational zero modes, eigenmodes are sorted in ascending order of their eigenvalues, i.e., $\omega^1 < \omega^2 < \cdots < \omega^{3N-3}$.
Then for a given mode, the indexes of particles are sorted in descending order of their norms, i.e., $|\boldsymbol{e}^k_1| > |\boldsymbol{e}^k_2| > \cdots > |\boldsymbol{e}^k_N|$. 
In order to focus only on the quasi-localized modes and exclude any effects due to hybridization with phonons, we picked up the vibrational modes located below the lowest frequency of phonons in our analysis~\cite{Lerner2016Statistics}
\footnote{
We first estimated the width of the band of the lowest frequency phonon by the same method as Ref.~\cite{Bouchbinder2018Universal}.
We then fitted the peak of the vDOS of the lowest frequency phonon to the Gaussian function and obtained the mean frequency $\omega_1$ and the standard deviation $\Delta \omega_1$ of the band. 
We finally estimated the width of the band to be $3\Delta \omega_1$, and we picked up only the vibrational modes that satisfy $\omega^k \in [0,\omega_1-3\Delta\omega_1]$ in our analysis.}.
In the main text, we denote the average over all the analyzed modes as $\left< \text{\textbullet} \right>_k$.

\begin{figure}[t]
\begin{center}
\includegraphics[width=0.4\textwidth]{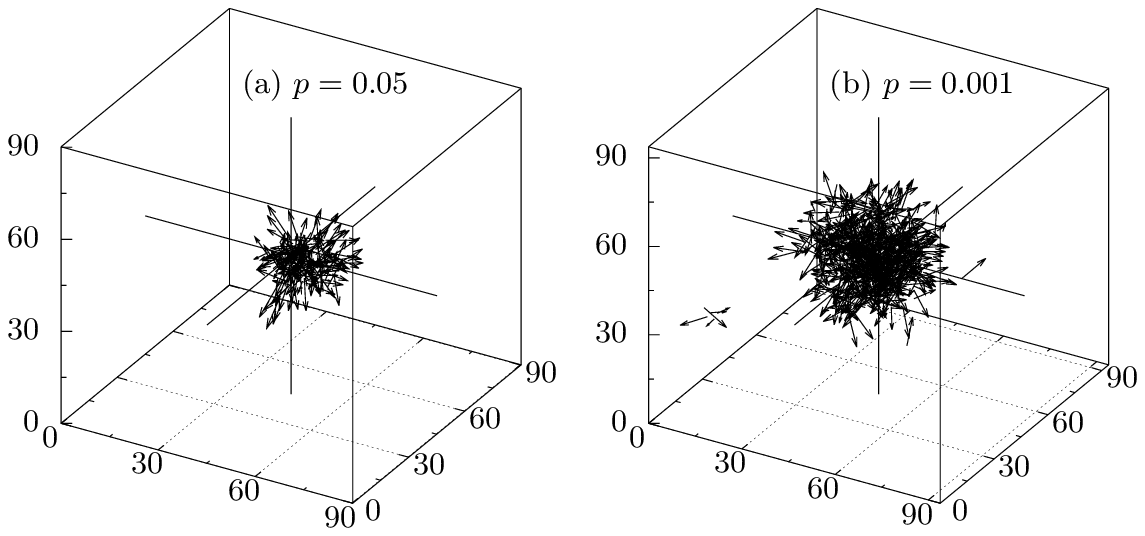}
\vspace*{0mm}
\caption{
Vibrational displacement field in the lowest-frequency mode at (a) the high ($p=0.05$) and (b) the low ($p=0.001$) pressures.
We show the particles' vibrational displacements (denoted by arrows) which are larger than 1\% of the largest one, and the particle with largest displacement is put at the center of the box.
}
\label{fig:fig1}
\end{center}
\end{figure}

\begin{figure}[t]
\begin{center}
\includegraphics[width=0.4\textwidth]{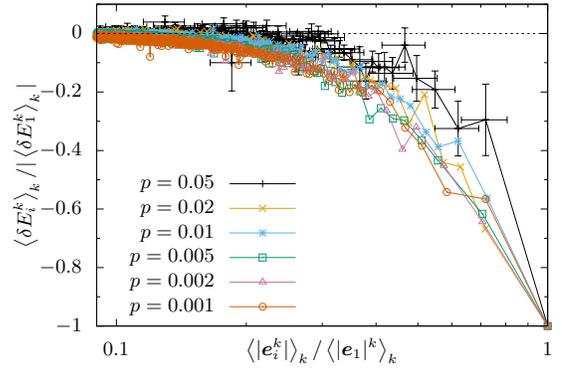}
\vspace*{0mm}
\caption{
The averaged energy versus the averaged norms of each particle.
Both axes are normalized by the values of the particle with largest displacement.
Error bars are shown only for pressure $p=0.05$, and the other errors are comparable.
The symbols are connected by lines to guide the eye.
}
\label{fig:fig2}
\end{center}
\end{figure}

\begin{figure}[t]
\begin{center}
\includegraphics[width=0.45\textwidth]{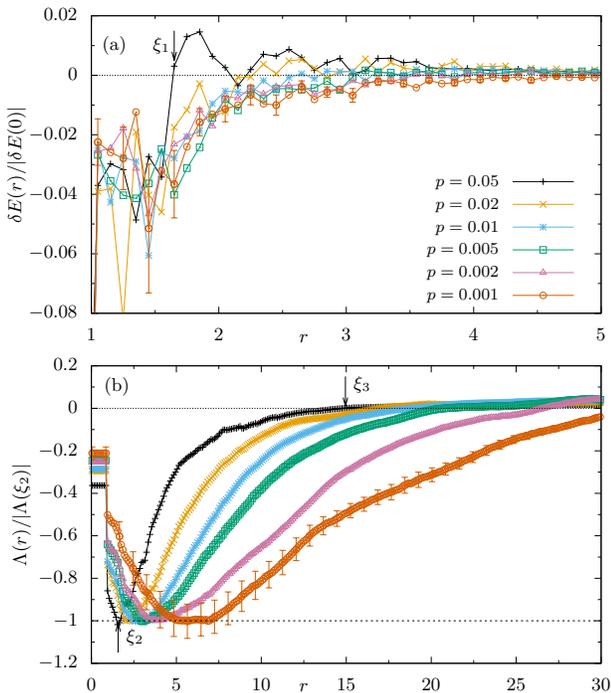}
\vspace*{0mm}
\caption{
(a) Radial energy distribution functions $\delta E(r) $ normalized by their values at origins for different pressures $p$.
We define the length $\xi_1$ where the functions firstly become positive and show it for $p=0.05$. 
(b) Integrated radial energy distribution function $\Lambda(r)$ normalized by their minimum values.
We also defined the length $\xi_2$ at which the functions become minimum and $\xi_3$ where the functions firstly become positive. 
For visibility in  (a) and (b) we show error bars only for $p=0.001$, other error bars are comparable.}
\label{fig:fig3}
\end{center}
\end{figure}

\begin{figure}[t]
\begin{center}
\includegraphics[width=0.45\textwidth]{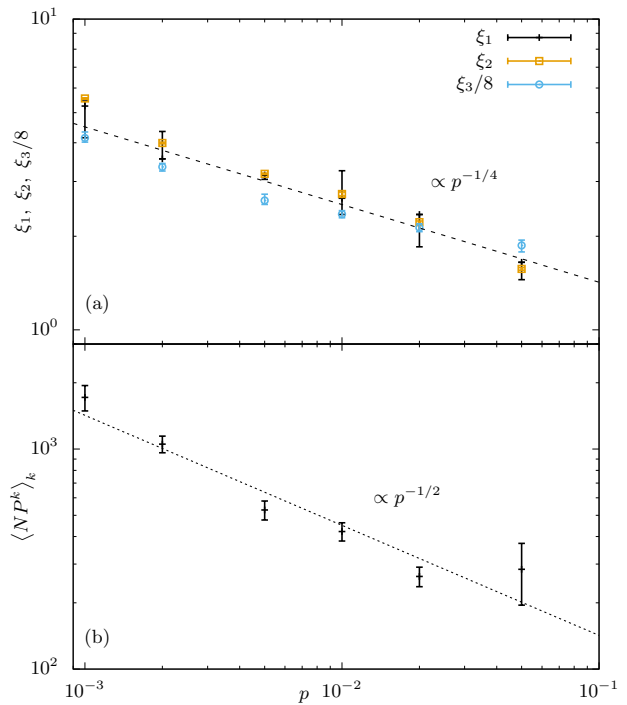}
\vspace*{0mm}
\caption{
(a) The pressure dependences of three lengths, $\xi_1, \xi_2, \xi_3$, defined in the main text and Fig.~\ref{fig:fig3}.
Since $\xi_3$ is much larger than the other two lengths, we present $\xi_3$ divided by $8$.
The dashed line indicates the power-law dependence of $\propto p^{-1/4}$.
(b) The pressure dependence of the volume.
The dotted line indicates the dependence of $\propto p^{-1/2}$.
}
\label{fig:fig4}
\end{center}
\end{figure}

{\it Results}.
Figure~\ref{fig:fig1} shows the visualization of the lowest-frequency modes of the systems at  high ($p=0.05$) and  low ($p=0.001$) pressures.
Each arrow indicates an eigenvector component $\boldsymbol{e}_i^k$,  only those larger than $1$\% of the largest one are shown. 
We observe that these modes present a core where the displacement is large and heterogeneous,
whose size appears to increases as pressure decreases. 

To characterize the motions of particles in these modes, we calculate the contribution $\delta E^k_i$ of particle $i$ to the energy of the mode $k$, which must satisfy $\lambda^k/2=\sum_i \delta E^k_i$.
It reads \cite{Alexander1998}: 
\begin{equation} \label{eq:particle energy}
\delta E^k_i = \frac{1}{4} \sum_{j\in\partial i}\left[\left(u_{ij}^{\parallel}\right)^2 - \frac{f_{ij}}{r_{ij}}\left(u_{ij}^{\perp}\right)^2\right],
\end{equation}
where $\partial i$ labels the set of particles interacting with particle $i$, $u_{ij}^{\parallel} = (\boldsymbol{e}_i^k-\boldsymbol{e}_j^k)\cdot\hat{\boldsymbol{r}}_{ij}$ is the relative displacement between $i$ and $j$ parallel to the bond $ij$ of direction $\hat{\boldsymbol{r}}_{ij}$, $u_{ij}^{\perp} = \sqrt{|u_{ij}^{\parallel}|^2 - (\left(\boldsymbol{e}_i^k-\boldsymbol{e}_j)\cdot\hat{\boldsymbol{r}}_{ij}\right)^2}$ is the perpendicular component of that relative displacement, and $f_{ij} = -\mathrm{d}\phi(r_{ij})/\mathrm{d}r$ is the contact force.
Note that $f_{ij}$ is always positive in the present system, and packings are called unstressed when setting $f_{ij} = 0$~\cite{Wyart2005Effects}.
Next we calculate the energy $\left< \delta E_i^k \right>_k$ of the $i$-th particle with largest displacement averaged on all the quasi-localized modes we obtain at a given pressure.
In Fig.~\ref{fig:fig2}, we plot $\left<\delta E_i^k\right>_k$ {\it v.s.}  the averaged norms $\left< |\boldsymbol{e}_i^k| \right>_k$. 
We find that the larger the norm, the lower the energy. In particular, particles in the core (particles with large norm) even have a {\it negative} energy.
This result implies that the perpendicular motion $u_{ij}^{\perp}$ is very dominant there, since it is the only  negative contribution to the energy  following Eq.~(\ref{eq:particle energy}), and $f_{ij}/r_{ij} \ll 1$ near jamming. 
Such a large perpendicular motion is a characteristic feature of anomalous modes ~\cite{Wyart2005Effects,Mizuno2016Elastic} and non-affine displacements under global deformations near jamming~\cite{Ellenbroek2009Jammed}, and of the transition between double well potentials~\cite{Heuer1996Collective}.

To study the spatial distribution of $\delta E_i^k$, we define the radial energy distribution function:
\begin{equation} \label{eq:RDFE}
\delta E(r) = \left<\frac{\sum_i\delta E_i^k\delta(r-r_i)}{\sum_i\delta(r-r_i)}\right>_k,
\end{equation}
where $r_i$ is the distance of the  particle $i$ from that with the lowest energy.
This function measures the average energy of particles at  distance $r$ from the center of the localized mode.
Fig.~\ref{fig:fig3}(a) shows $\delta E(r)$ for  different $p$.  For the moment, we focus on data that are far away from jamming, corresponding to $p=0.05$ (black line).
We observe  that $\delta E(r)$ is negative up to some length scale we denote as $\xi_1$, here  $\xi_1 \approx 1.5$. For $r \gtrsim \xi_1$, $\delta E(r)$ is a positive quantity
and decays rapidly with distance as expected from the decay of the displacements themselves. $\xi_1$ thus characterizes the size of the unstable core of the localized modes, which is stabilized 
by its far field components corresponding to $r>\xi_1$.


We then calculate the integrated radial energy distribution function defined as: 
\begin{equation} \label{eq:IRDFE}
\Lambda(r) = \left<\sum_{r_i\leq r}\delta E_i^k\right>_k. 
\end{equation}
$\Lambda(r)$ corresponds to the average energy the localized modes would have if the system was cut at a distance $r$ from the center of the mode. 
Obviously, $\lim_{r\to\infty}\Lambda(r)=\left<\lambda^k \right>_k/2$. 
There is a direct link between $\Lambda(r)$  and $\delta E(r)$:  
\begin{equation}
\label{Integral}
\Lambda(r) \approx \int d\boldsymbol{r}'\rho G(r')\delta E(r')
\end{equation}
where $\rho$ is the number density, and $G(r)$ is the radial distribution function~
\footnote{This is an approximate relation because the normalization by the number of particles is done before the average for $\delta E(r)$ , whereas not for $\Lambda(r)$.}.
$\Lambda(r)$ are shown for different pressures in Fig.~\ref{fig:fig3}(b).
Again, we focus on $p=0.05$ for the moment (black line). From Eq.~(\ref{Integral}), it is clear that the negativity of $\delta E(r)$
at small distances results in the negativity of $\Lambda(r)$ at small $r$, which must display a minimum at a distance $\xi_1$ defined above. 
For  $r>\xi_1$, $\Lambda(r)$ gradually increases and becomes positive at a distance  we denote $\xi_3$. Here  $\xi_3\approx 15$,
which is ten-fold larger than the core size $\xi_1$.   In practical terms, this result implies that even far from jamming, cutting the system around a localized mode
at rather large distances $r<15$ (and imposing external forces at the particles at the boundary  to maintain force balance) would not lead to a stable system:
the localized mode would still be unstable, and rearrangements would necessarily occur. The emerging physical picture for quasi-localized modes is that of a core which
is passed a structural buckling instability \cite{Alexander1998,Wyart2005Effects}, but which is stabilized by the surrounding elastic medium. This situation is similar to confined thin sheets where buckling can be prevented 
by adhering the system to a surrounding stabilizing elastic medium \cite{Cerda03}.

We  now study how the architectures of the quasi-localized modes depend on the proximity to jamming. 
We consider three different lengths from the observables introduced above. 
We recall that $\xi_1$ is defined as the length where $\delta E(r)$ becomes positive. We define  $\xi_2$ has the length where $\Lambda(r)$ reaches a minimum,
which must satisfy $\xi_1 \approx \xi_2$ due to Eq.~(\ref{Integral}). Lastly,  $\xi_3$ is smallest $r$ for which $\Lambda(r)$ becomes  positive.
$\xi_1$, $\xi_2$ and $\xi_3$ are indicated in Figs.~\ref{fig:fig3}(a) and \ref{fig:fig3}(b) by arrows for $p=0.05$. 
The pressure dependences of $\xi_1,\xi_2,\xi_3$ is shown in Fig.~\ref{fig:fig4}(a), which supports  the following power-law dependence:
\begin{equation}
\label{length}
\xi_1, \xi_2, \xi_3 \propto p^{-1/4}.
\end{equation}
Therefore, the quasi-localized modes become more extended as $p\rightarrow 0$, and their characteristic length scale diverges at jamming.

Another characterisation of these modes is their participation ratio 
$P^k = \frac{1}{N}\left[\sum_i\left(\boldsymbol{e}_i^k\cdot\boldsymbol{e}_i^k\right)^2\right]^{-1}$. 
The quantity $ NP^k $ is an estimate of the number of particles involved in the mode $k$ ~\cite{Mazzacurati1996Low,Taraskin1999Anharmonicity,Schober2004Size}.
We define  the average volume of the localized modes as  $ V\equiv\left< NP^k \right>_k$, whose dependence on $p$ is shown in Fig.~\ref{fig:fig4}(b).
Once again we find a singular behavior near jamming, consistent with: 
\begin{equation}
\label{volume}
V \propto p^{-1/2}.
\end{equation}

{\it Discussion}.
The scaling results Eqs.~(\ref{length}) and (\ref{volume}) support that the quasi-localized modes are  the anomalous modes responsible for the Boson peak in these systems, whose properties we now recall. Near  jamming,  the density of vibrational modes exhibits a flat spectrum  $g(\omega)\sim\mathrm{\omega^0}$ at frequencies $\omega > \omega_\ast$~\cite{O'hern2003Jamming,Silbert2005Vibrations}, where $\omega_\ast \propto p^{1/2}$. 
Anomalous modes at $\omega_\ast$ are spatially extended, but can be characterized by finite correlation length which diverges at the jamming transition as $\ell_c \propto p^{-1/4}$~\cite{Silbert2005Vibrations,Silbert2009Normal,Ikeda2013}, 
a length scale that also characterizes the response to a local perturbation ~\cite{During2013Phonon,Lerner2014Breakdown,Karimi2015Elasticity}. These results can be derived via effective medium calculations \cite{Wyart2010Scaling,DeGiuli2014Effects}. Anomalous modes at $\omega_\ast$ and the response to a local perturbation are also tied together by a recent variational argument~\cite{Yan2016On} showing that the later can be used as building blocks to reconstruct the former. These building blocks can be localized on a length scale $l_c$ and on a characteristic volume $V \propto p^{-1/2}$~\cite{Yan2016On} without affecting significantly their frequency scale. ($V \propto p^{-1/2}$ differs from $l_c^d$ where $d$ is the spatial dimension due to the algebraic decay of the mode magnitude in space). The architecture we discovered for quasi-localized modes is thus fully consistent with that of the building blocks of anomalous modes. 

This correspondence can be used to explain the existence of the quasi-localized modes and to predict how their density depend on the distance to jamming. In the absence of pre-stress (obtained by dropping the second term in Eq.~(\ref{eq:particle energy})), there are no anomalous modes at frequencies $\omega<\omega_\ast$, a frequency beyond which they suddenly appear and their density becomes large~\cite{Wyart2005Effects}. Due to this large density, it is plausible that these modes hybridize and are thus extended. 
In the stressed system, the energy of anomalous modes  decreases approximately by $-p$ due to the second term in Eq.~(\ref{eq:particle energy}) ~\cite{Wyart2005Effects}. As a result, anomalous modes populate the entire frequency range $0<\omega<\omega_\ast$, an effect coined marginal stability.  
This effect lifts the degeneracy of the anomalous modes, which then become quasi-localized on the characteristic length scale of the building blocks that constitute them. 
This view  is consistent with the finding that quasi-localized modes are mostly apparent in the stressed system, and disappear when pre-stress is removed ~\cite{Mizuno2017Continuum,Lerner2017Frustration}. 
The integrated density of anomalous modes in this frequency range is of order $\omega_\ast$. This scaling, implied by the flat density of anomalous modes, simply states that there is one anomalous modes in this frequency range every volume $V$. Let us assume that a finite fraction of these modes become quasi-localized. From general arguments \cite{Gurarie03}, we know that their density must follow $g_{\mathrm{loc}}(\omega) \sim c(p) \omega^4$. Requiring that:
\begin{equation}
\int_0^{\omega_\ast} g_{\mathrm{loc}}(\omega) d\omega \sim \omega_\ast \sim \frac{1}{V}
\end{equation}
fixes $c(p)\sim V^{-1} \omega_\ast^{-5} \propto p^{-2}$, as  indeed found numerically~\cite{Mizuno2017Continuum}. Our approach thus rationalises why the density of quasi-localized modes exploses near jamming.  

Overall, our work supports that quasi-localized modes correspond to the anomalous modes known to control the boson peak in finite-range interacting systems. 
Although this correspondence is most stringently tested near jamming where both objects display singular properties, we expect it 
to hold true away from jamming as well. If so, our conclusion should hold in  Lennard-Jones glasses, where the boson peak can also be interpreted in terms of the distance to a jamming transition (that cannot vanish however due to long-range interactions) ~\cite{Xu2007Excess},  but also in chalcogenide glasses and silica where jamming corresponds to the point where the covalent network becomes rigid \cite{Phillips1979,DeGiuli2014Effects}.

\begin{acknowledgments}
We thank H. Yoshino, E. Lerner, L. Manning, M. Popovic, T. DeGeus and W. Ji  for discussions. This work was supported by JSPS KAKENHI Grant Numbers JP17K14369, JP17H04853, JP16H04034, the Swiss National Science Foundation under Grant No. 200021-165509 and the Simons Foundation (Grant \#454953 Wyart).
The numerical calculations were partly performed in Research Center for Computational Science, Okazaki, Japan. 
\end{acknowledgments}

\bibliographystyle{apsrev4-1}
\bibliography{draft}

\end{document}